\documentclass[a4paper,11pt]{article}
\pdfoutput=1 

\usepackage{jcappub} 

\usepackage[T1]{fontenc} 
\usepackage[utf8]{inputenc}
\usepackage{subfig}

\title{\boldmath Light Dark Matter through Assisted Annihilation}

\author[a]{Ujjal Kumar Dey,}
\author[b]{Tarak Nath Maity,}
\author[b]{Tirtha Sankar Ray}

\affiliation[a]{Centre for Theoretical Studies, Indian Institute of Technology, Kharagpur 721302, India}
\affiliation[b]{Department of Physics, Indian Institute of Technology, Kharagpur 721302, India}

\emailAdd{ujjal@cts.iitkgp.ernet.in}
\emailAdd{tarak.maity.physics@gmail.com}
\emailAdd{tirthasankar.ray@gmail.com}

\abstract{In this paper we investigate light dark matter scenarios where annihilation to Standard Model particles at tree-level is kinematically forbidden. In such cases annihilation can be aided by massive Standard Model-like species, called {\it assisters}, in the initial state that enhances the available phase space opening up novel tree-level processes. We investigate the feasibility of such non-standard {\it assisted annihilation} processes to reproduce the observed relic density of dark matter. We present a simple scalar dark matter - scalar assister model where this is realised. We find that if the dark matter and assister are relatively degenerate the required relic density can be achieved for a keV-MeV scale dark matter. We briefly discuss the cosmological constraints on such dark matter scenarios.
}


\begin{document}
\maketitle
\flushbottom

\section{Introduction}
\label{sec:intro}
A multitude of astrophysical and cosmological observations, encompassing scales ranging from galactic to hundreds of megaparsecs, indicate that about 85\% of the total mass of the Universe consists of dark matter (DM) which leaves its footprints in those observations through gravitational effects. The exact nature, origin, mass, composition of DM is yet to be uncovered and consequently it does not fit in the Standard Model (SM) of particle physics and demands for physics beyond the Standard Model (BSM); see e.g.,~\cite{Jungman:1995df, Munoz:2003gx, Bertone:2004pz, Bergstrom:2012fi}, for a review. An attractive candidate for DM is the so-called weakly interacting massive particles (WIMP), various incarnations of which arise naturally in many BSM scenarios, e.g., $R$-parity conserving supersymmetry, KK-parity conserving universal extra dimensional models, inert doublet model etc. In all of these scenarios, an unbroken symmetry in the theory accounts for the stability of the DM particle. A discrete $\mathbb{Z}_{2}$ symmetry provides the minimal realisable framework for this\footnote{Recently, in the context of semi-annihilation, self-interaction and other mechanisms, various models using higher stabilising symmetry e.g., $\mathbb{Z}_{3}$ (and generalised $\mathbb{Z}_{N}$) have been constructed, see e.g.,~\cite{D'Eramo:2010ep, Belanger:2012vp, Ivanov:2012hc, Belanger:2012zr, Ko:2014nha, Belanger:2014bga, Aoki:2014cja, Bernal:2015bla, Choi:2015bya, Cai:2015tam, Ding:2016wbd, Karam:2016rsz, Karam:2015jta}.}. The freeze-out of the dark matter is usually facilitated by a $2\to 2$ annihilation of DM to SM states.  
The allowed range of mass for DM candidates spans over large orders of magnitude from sub-eV \cite{Hui:2016ltb} to $~{\mathcal{O}}(100)$ TeV~\cite{Griest:1989wd}. Standard WIMP scenarios considers  $\sim$ GeV mass cold dark matter (CDM) candidates with typical weak scale interactions that successfully explains the Universe at large scales. However, in the galactic scales CDM faces difficulties in explaining the so-called {\it missing satellite}, {\it too-big-to-fail} and the {\it core vs. cusp} problems arising from the mismatch between $N$-body simulations and astrophysical observations~\cite{Weinberg:2013aya}. It has been shown by cosmological simulations that these problems can be addressed by sub-GeV self-interacting dark matter~\cite{Spergel:1999mh, 2012MNRAS.423.3740V, Rocha:2012jg,  Peter:2012jh, Hochberg:2014dra}. 
Under the auspices of these observations, the preferred DM mass can be so light that the usual paradigm of $2\to 2$ annihilation of DM to SM particles may be phase space suppressed or even kinematically forbidden and thus the freeze-out process may stall. Note that annihilation to massless SM degrees of freedom can not be technically forbidden. However, we are interested in scenarios where such annihilation is highly suppressed and arise, for example, at higher loops to have any significant numerical effect. To facilitate the successful freeze-out in such scenarios, we propose the idea of {\it assisted annihilation}, where the light DM particle thermalises through the assistance of other particle(s) which we call {\it assister(s)}. These massive assisters, by taking part in the initial state, help to surmount the phase space barrier for the annihilation of the DM particles. The assisters, not being charged under any stabilizing symmetry, have interactions with the SM states such that they can decay to SM particles. Thus, we can consider the  assisters to be SM-like, in the sense that they follow the similar thermal history as the SM particles. The assisted annihilation framework as described above is clearly  different from the well-known coannihilation scenario where the particle that coannihilates with the DM is also charged under the same stabilising symmetry. The assisted annihilation is also different from the ``assisted freeze-out'', considered in~\cite{Belanger:2011ww}, having a two component dark matter where the thermalisation of one DM particle occurs via the assistance of the other component.
Apart from the relic density considerations, there are also state-of-the-art experiments that are looking for the direct detection of DM via the DM-nucleon scattering. The ever-increasing precision in the measurement of both the spin-dependent and spin-independent scattering cross section severely constrain the parameter space for DM models~\cite{Aprile:2016swn, Akerib:2016lao, Savage:2015xta, Fu:2016ega}. Satellite-based experiments, e.g., Fermi-LAT, AMS-02 etc., searching for signals of DM annihilation in the local neighbourhood provide an handle for indirect detection of DM and constrains several standard scenarios~\cite{Lavalle:2012ef, Conrad:2014tla, Cirelli:2015gux}. However, in the case of assisted annihilation/semi-annihilation the annihilation topologies are such that it is difficult to probe the scenario either in direct or indirect detection experiments.    
In the present paper, we construct a simplistic scalar model to demonstrate the idea of assisted annihilation. We find viable DM candidates in the keV-MeV mass range along with relatively degenerate assisters. We show that the $4\to 2$ assisted annihilation can give rise to a keV scale DM  whereas MeV scale DM can be obtained when $3\to 2$ processes are operative. In the keV case we demonstrate that the DM can very well be cold and satisfy the observed relic density, at the same time maintaining the Ly-$\alpha$ constraints. Moreover, it is possible to address the small scale structure formation issues by tuning the self-coupling of the keV scale DM particles, without hampering the proposed assisted annihilation. Similarly, for the case of MeV scale DM we find it to be cold in nature and the parameter space allowed by relic density considerations is also in consonance with constraints from large scale structure formations. These models provide a proof of principle for the proposed assisted annihilation mechanism while more involved and possibly more realistic versions of these models may be realized in nature. We briefly outline a scenario with a scalar DM and a vector boson assister.
The rest of the paper is organised as follows. In the next section we present the general idea and the analytical details of the assisted annihilation framework. In section~\ref{sec:scalarAsst} we construct a minimal scalar model for the assisted annihilation and analyse the parameter space allowed by relic density and other constraints. We also give a brief sketch of a model with vector boson as assisters, without resorting to rigorous parameter space scanning in section~\ref{sec:vecAsst}. Finally we summarise and conclude in section~\ref{sec:concl}. In the appendix we spell out some of the details of the calculation of amplitudes in the non-relativistic limit and the semi-analytic expression for the freeze-out temperature.

\section{Assisted annihilation framework} 
\label{sec:asstann}
\begin{figure}[t]
\begin{center}
\subfloat[\label{sf:asstannhln}]{
\includegraphics[scale=0.6]{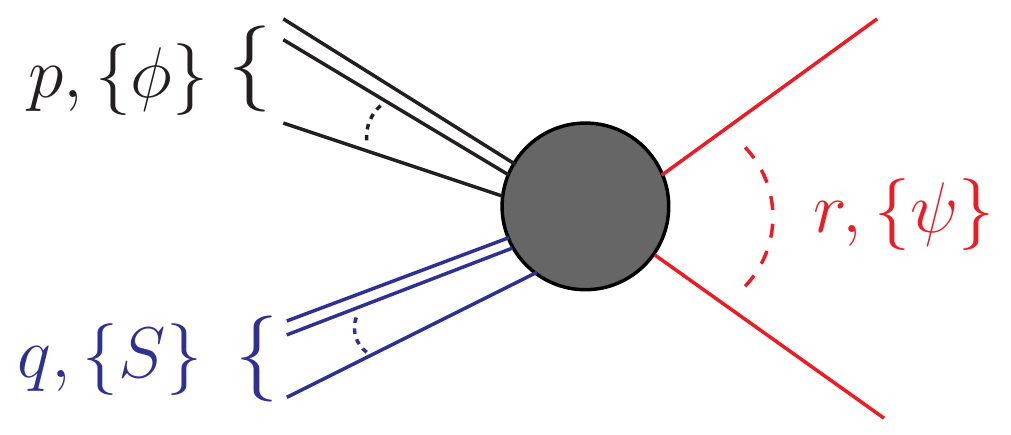}}~~~~
\subfloat[\label{sf:asstSemiAnnhln}]{
\includegraphics[scale=0.6]{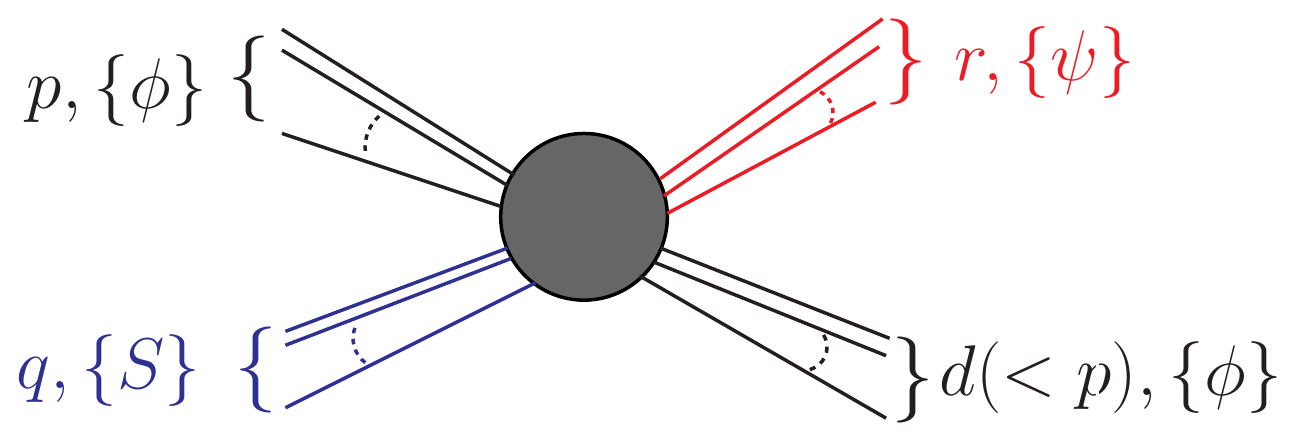}}
\caption{Schematic diagrams representing the (a) assisted annihilation and (b) assisted semi-annihilation.}
\label{fig:annhln}
\end{center}
\end{figure}

\begin{figure}[t]
\begin{center}
\subfloat[\label{sf:nbolt}]{
\includegraphics[scale=0.75]{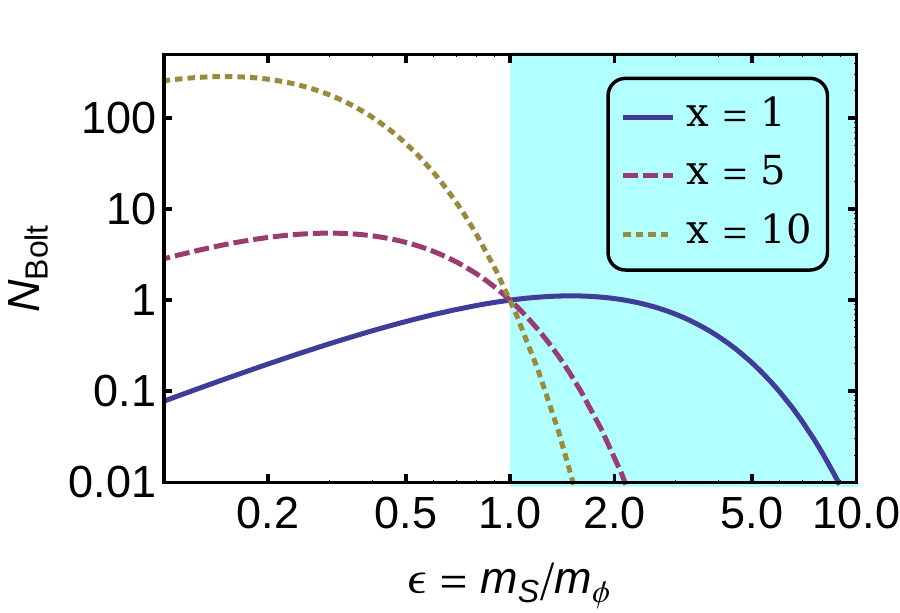}}~~~~
\subfloat[\label{sf:freez}]{
\includegraphics[scale=0.7]{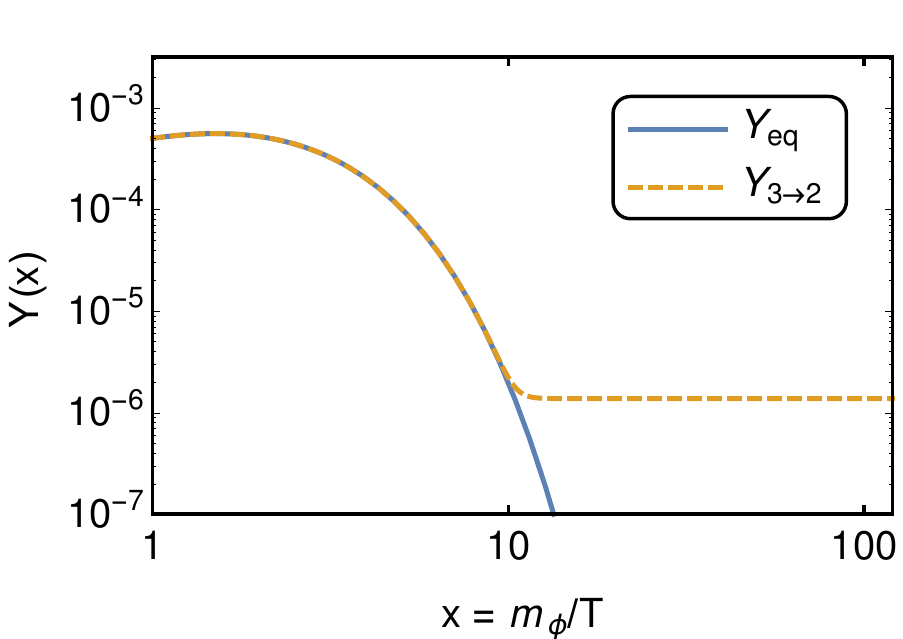}}
\caption{(a) Boltzmann factor $N_{\rm Bolt}$ as a function of mass ratio between assister and DM mass for various $x$ for $p = 2$, $q=1$ and $r = 2$. (b) Freeze out curve for an $\mathcal{O}$(MeV) DM particle $\phi$. For this we have considered the assisted annihilation of $\phi$ via the process $\phi \phi S \to \psi \psi$ with a realistic thermally averaged cross section of $1.82 \times 10^{12}$ GeV$^{-5}$.}
\label{fig:nboltfreez}
\end{center}
\end{figure}

In this section we lay out the generic framework for assisted annihilation. The basic ingredients of this setup is the DM species, $\{\phi\}$ and the assisters, $\{S\}$. For the time being the nature of $\phi$ and $S$ is unspecified and in principle, they can be scalars, fermions or gauge bosons. For simplicity we assume the stability of the DM particle, $\phi$ is determined by a discrete $\mathbb{Z}_{2}$ symmetry. The assister, $S$ is assumed to be SM-like, in the sense that they can decay to SM states and thus maintain the thermal equilibrium with the visible sector. We assume that the equilibrium number density of the assisters follow the usual evolution of the SM particle species. 
Assisted annihilation becomes relevant in the region of parameter space when the DM is lighter than the assisters or other SM particles that it couples to. In such a case the traditional $2\to 2$ annihilation of the DM particles might be kinematically forbidden at tree-level. However, in such a scenario the assisters can come up in the initial state and {\it assist} to open up new channels (e.g., $3\to 2$, $4\to 2$ etc.) for the annihilation of DM. In figure~\ref{sf:asstannhln}, we show a typical assisted annihilation process. This involves the annihilation of $p$ (even integer) number of dark matter species with $q$ asisters. The final state contains $r$ number of particles that can either be assisters or any other SM fields. It is worth-mentioning that the assisted annihilation is different from the coannihilation where the coannihilating species are charged under the same stabilising symmetry. This should be contrasted with the present setup where the assisters are not charged under the DM stabilising symmetry and therefore can be an SM state itself.
For a generic process like, $\underbrace{\phi, \ldots, \phi}_{p}, \underbrace{S, \ldots, S}_{q} \to \underbrace{\psi, \ldots, \psi}_{r}$, where $\{\psi\}$ are generic SM-like (i.e., SM or assister) final states, the general form of the Boltzmann equation to determine the number density of the species $\phi$ is given by,
\begin{align}
\frac{dn_{\phi}}{dt} + 3Hn_{\phi} = - \int 
          \prod_{\substack{i = 1,\ldots,p,\\1,\ldots,q}}d\Pi_{i}
          \int & \prod_{j = 1,\ldots,r}d\Pi_{j} (2\pi)^{4}
           \delta\bigg(
           \sum_{\substack{i = 1,\ldots,p,\\1,\ldots,q}}p_{i}
           -\sum_{j = 1,\ldots,r}p_{j}\bigg) \notag \\
           & \times |\mathcal{M}|^{2}\bigg(
           \prod_{\substack{i = 1,\ldots,p,\\1,\ldots,q}}f_{i}
           -\prod_{j = 1,\ldots,r}f_{j}\bigg),
\label{eq:boltzn}
\end{align}
where $H$ is the Hubble parameter which is given as, $ H(m_{\phi}) = \sqrt{\pi^{2}g_{\ast}/90}(m_{\phi}^{2}/M_{\rm Pl})$, with $g_{\ast}$ being the effective degrees of freedom and $M_{\rm Pl}$ is the Planck scale, and $f_{X}$ is the distribution function for the species $X$. In writing this equation the standard assumptions of $T$ (or $CP$) invariance and $1+f_{X} \simeq 1$ are made. We can rewrite eq.~\ref{eq:boltzn} in terms of the dimensionless parameters $x \equiv m/T$ and comoving number density $Y \equiv n/s$ as follows,
\begin{subequations}
\begin{align}
\label{seq:boltzYY}
 \frac{dY_{\phi}}{dx} &= -\frac{x s^{p+q-1}}{H(m_{\phi})}
                        N_{\rm Bolt}
                        \langle \sigma v^{p+q-1}\rangle
                        \left[\left(Y_{\phi}\right)^{p}
                        \left(Y_{\phi}^{\rm eq}
                        \right)^{q} - \left(Y_{\phi}^{\rm eq}\right)^{p+q}
                        \right], \\
\label{seq:nbolt}
N_{\rm Bolt} &= e^{qx(1-\epsilon)}\epsilon^{3q/2},       
\end{align}
\label{eq:boltzy}
\end{subequations}
where the entropy density, $ s = 2 \pi^{2} g_{\ast}T^{3}/45$, $\epsilon = m_{S}/m_{\phi}$ and $v$ represents the average relative velocity of the colliding particles in the initial state. The statistical factor $N_{\rm Bolt}$ modulates the annihilation cross section of the process.  Physically it captures the feature that assisted annihilation can continue as long as the number density of both the DM states and the assisters are appreciable in the early universe. If the assisters disappear much before the DM, this mechanism essentially stalls and cannot drive  a successful freeze-out. This essentially dictates that the assisters while being heavier than the DM, needs to be reasonably degenerate with the DM masses which is reminiscent of the degeneracy required in co-annihilation processes that originate from similar considerations. The numerical impact of the factor $N_{\rm Bolt}$ is shown in figure~\ref{sf:nbolt} for typical values of various parameters in the theory.  In the region of interest for $\epsilon > 1,$ the factor rapidly reduces to small values suppressing the assisted annihilation process\footnote{The case $\epsilon < 1$ corresponds to the region of parameter space where the usual $2\to 2$ annihilation becomes operative.}. In what follows we will restrict ourselves to $\epsilon \sim {\mathcal{O}}(1)$  so that the suppression from this statistical factor is not crippling. In passing, we note that a more rigorous approach includes the solution of coupled Boltzmann equation for the DM as well as assisters including the decay of the assisters to SM states. However, the relative degeneracy of the assisters and the DM states keeps it a viable assumption that when the DM decouples the assisters are still in non-relativistic thermal equilibrium.
We will confine ourselves to annihilation topologies with at most two species in the final state for simplicity. The thermally averaged cross section for a $p+q\to 2$ annihilation dimensionally scales as,
\begin{align} \label{eq:dim}
\left[\sigma_{p+q\to 2}v^{p+q-1}\right] = 
                \left[M^{-3(p+q)+4}\right].
\end{align}
From this, it is evident that to have appreciable annihilation we require the least number of initial state particles that allows kinematically viable assisted annihilation in a given model.  In the present work we confine ourselves to the examples of assisted annihilation and assisted semi-annihilation via $3\to 2$ and $4\to 2$ processes. Therefore, in our case, $p+q=3,4$ and $r=2$ in figure \ref{sf:asstannhln}. Further,  if a $3 \to 2$ process is available the contribution of $4 \to 2$ becomes numerically negligible due to the large mass suppression as given in eq. \ref{eq:dim}. The general expression for the thermally averaged cross section in the non-relativistic limit for a $p+q \to 2$ process, with initial state particle masses $m_{i}~(i=1,\ldots,(p+q))$ and final state particle masses $m_{f_{k}}~(k=1,2)$, is given by,
\begin{align}
\langle \sigma v^{p+q-1} \rangle = 
                     \frac{1}{4 \pi}\prod_{i=1,\ldots,(p+q)}
                     \frac{1}{2 m_i M^{2}} \sqrt{M^{4}-2
                      M^2(m^2_{f_{1}}+m^2_{f_{2}})+{m^{\prime}}^{4}}
                      \lvert \mathcal{M}_{(p+q) \to 2}
                      \rvert^{2}_{\rm NR},
\end{align} 
where, $M = \sum_{i=1}^{p+q} m_i$ and ${m^{\prime}}^{2} = m^{2}_{f_{1}}-m^2_{f_{2}}$. 
In figure~\ref{sf:freez}, we show the freeze-out occurring in a typical $3\to 2$ process with realistic cross section. As one can see, this leads to a freeze-out similar to a cold DM scenario.

Similar to the assisted annihilation one can have assisted semi-annihilation as shown in figure~\ref{sf:asstSemiAnnhln}. One can easily set up the corresponding Boltzmann equation by following the general prescription of eqs.~\ref{eq:boltzn} and \ref{eq:boltzy}.
As a proof of principle for assisted annihilation/semi-annihilation we discuss, in the next section, a calculable minimal model for a scalar DM and scalar assister, that can satisfy the observed relic abundance through the assisted annihilation process while maintaining other structure formation constraints. We show that this naturally lead to keV-MeV scale light dark matter scenarios.  

\section{A minimal scalar model} 
\label{sec:scalarAsst}
In this section we present a simple scalar DM and scalar assister scenario that annihilate through assisted annihilation in certain region of parameter space to give the observed DM relic abundance~\cite{Ade:2013zuv}. This opens up hitherto forbidden windows of light DM in the parameter space of these models.
As mentioned earlier, we focus on the assisted annihilation processes via $4\to 2$ and $3\to 2$ processes. To maintain the relic density constraints via annihilation processes where the number of initial state particles are more than two, the DM candidates have to have masses in the sub-GeV range. Since the assisters are more or less degenerate in mass with the DM, their masses will also be in the same ballpark. We find that depending on whether the assisted annihilation processes are $4\to 2$ or $3\to 2$ processes the spectrum can be in keV and MeV range respectively. Below we illustrate each case separately.
\begin{figure}[t]
\begin{center}
\includegraphics[scale=0.5]{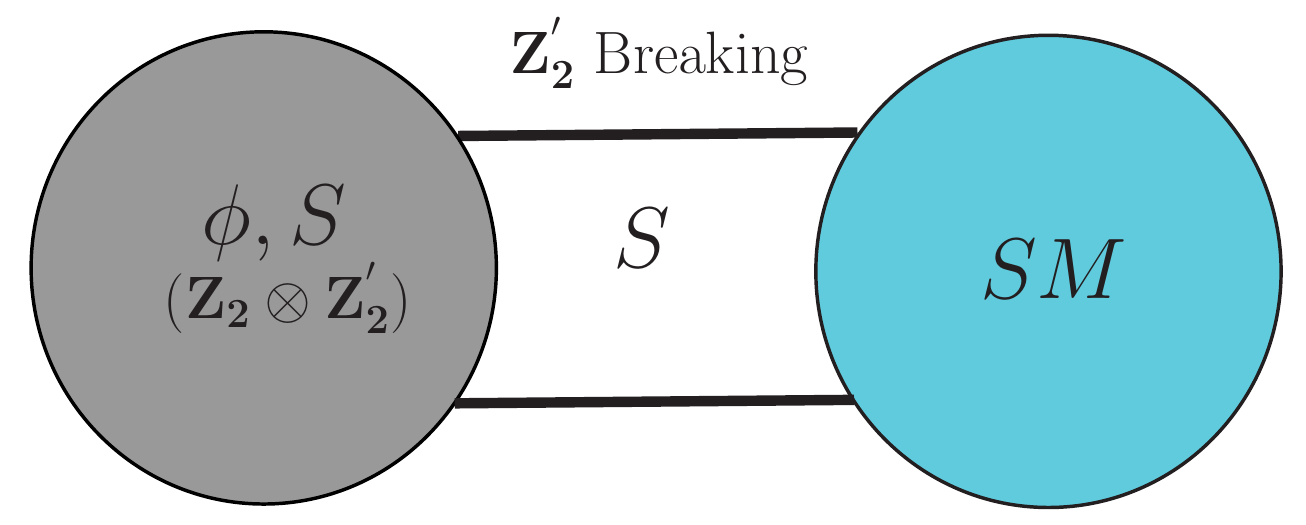}
\caption{Schematic diagram representing the dark sector and the visible (SM) sector. The breaking of $\mathbb{Z}_{2}^{\prime}$ results in the connection between the dark sector and the visible sector.}
\label{fig:cartoon}
\end{center}
\end{figure}
%
\subsection{keV scale dark matter}
Consider a dark sector consisting of a scalar dark matter  $\phi$ and one  scalar assister $S.$ Assume that the dark sector has an apparent $\mathbb{Z}_{2}\otimes \mathbb{Z}_{2}^{\prime}$ symmetry. The fields $\phi$ and $S$ are odd under $\mathbb{Z}_{2}$ and $\mathbb{Z}_{2}^{\prime}$, respectively. While, all the SM fields are even under both $\mathbb{Z}_{2}$ and $\mathbb{Z}_{2}^{\prime}$. The relevant Lagrangian for the dark sector includes,
\begin{align}
\label{eq:lag}
\mathcal{L}^{\rm dark} \supset \frac12 m_{\phi}^{2} \phi^{2} + 
              \frac12 m_{S}^{2} S^{2} + 
              \frac{\lambda_{\phi S}}{4}\phi^{2}S^{2} + 
              \frac{\lambda_{\phi}}{4!} \phi^{4} + 
              \frac{\lambda_{S}}{4!}S^{4}.
\end{align}
The $\mathbb{Z}_{2}^{\prime}$ is explicitly broken through the interaction of $S$ to SM states leaving an unbroken $\mathbb{Z}_2$ that stabilizes the DM. This is schematically shown in figure~\ref{fig:cartoon}. Assuming the assisters are leptophilic the $\mathbb{Z}_{2}^{\prime}$ breaking Lagrangian  includes,
\begin{align}
\label{eq:lagz2brk}
\mathcal{L}_{\not{\mathbb{Z}_{2}^{\prime}}}^{\rm med-SM} 
                        \supset 
       \lambda_{i}S\bar{\ell_{i}}\ell_{i}, 
\end{align}
where $i = \{ e,\mu,\tau\}$ is the flavour index. Depending on the mass the assisters $S$ can decay to SM states either directly or through loop of leptons $\ell_{i}$. Note that $\lambda_{i}$ is suppressed by the precision measurements of anomalous magnetic moments $(g-2)_{\ell_{i}}$. The contribution of neutral scalar $S$, in the $(g-2)_{\ell_{i}}$ is given by~\cite{Jackiw:1972jz, Jegerlehner:2009ry},
\begin{align}
\Delta a_{\ell_{i}} \simeq 
              \frac{\lambda_{i}^{2} 
              m_{\ell_{i}}^{2}}{4\pi^{2}}
              \int_{0}^{1}dx 
              \frac{(2-x)x^{2}}{m_{\ell_{i}}^{2}x^{2} 
              + (1-x)m_{S}^{2}}.
\end{align}
The experimental values of $\Delta a_{\ell_{i}}$~\cite{Olive:2016xmw} put constraints on $\lambda_{i} < \{ 10^{-6},10^{-4},0.3 \}$ for $\{e,\mu,\tau\}$ respectively. Thus appreciable coupling is only allowed to the third generation of leptons. 
\begin{figure}[!htbp]
\begin{center}
\includegraphics[scale=0.36]{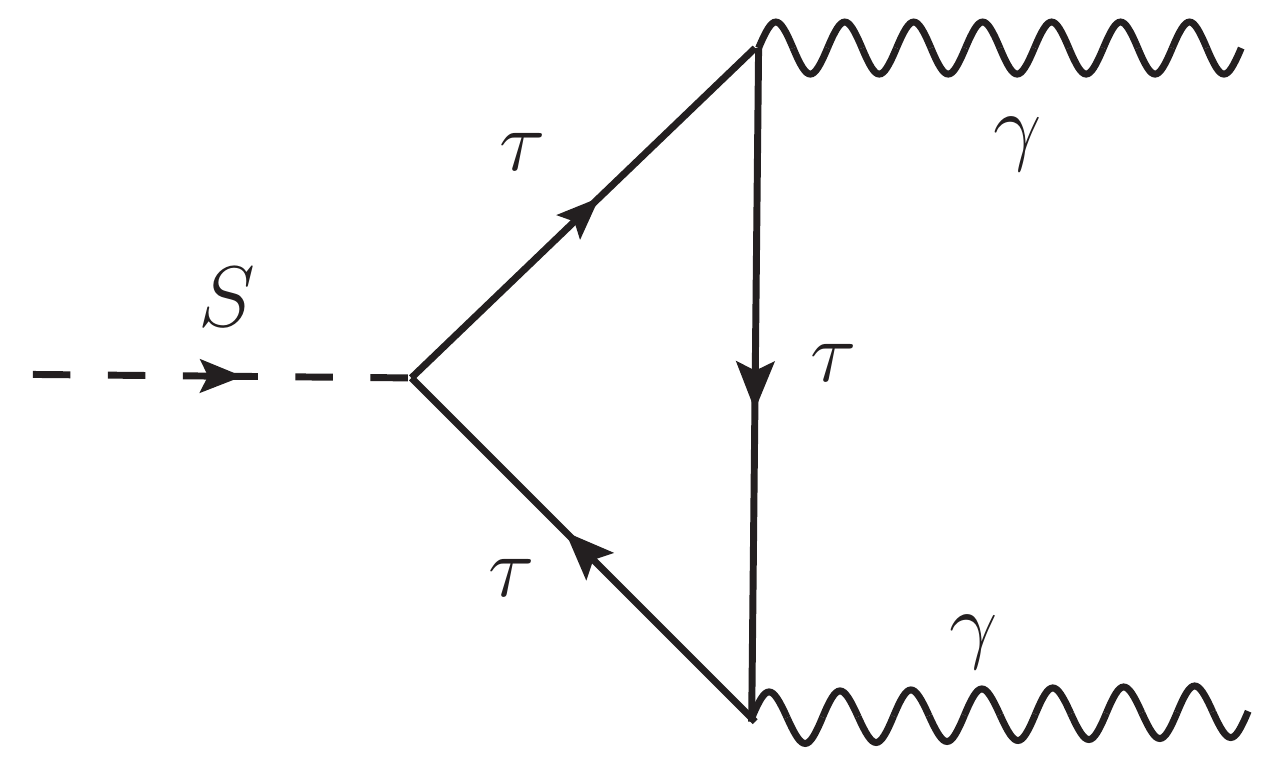}~~~~~
\caption{Feynman diagrams for loop-induced decay of $S$.}
\label{fig:syy}
\end{center}
\end{figure} 
\begin{figure}[!htbp]
\begin{center}
\subfloat[\label{sf:4to2_LTmSby2_mS15keV}]{
\includegraphics[scale=0.35]{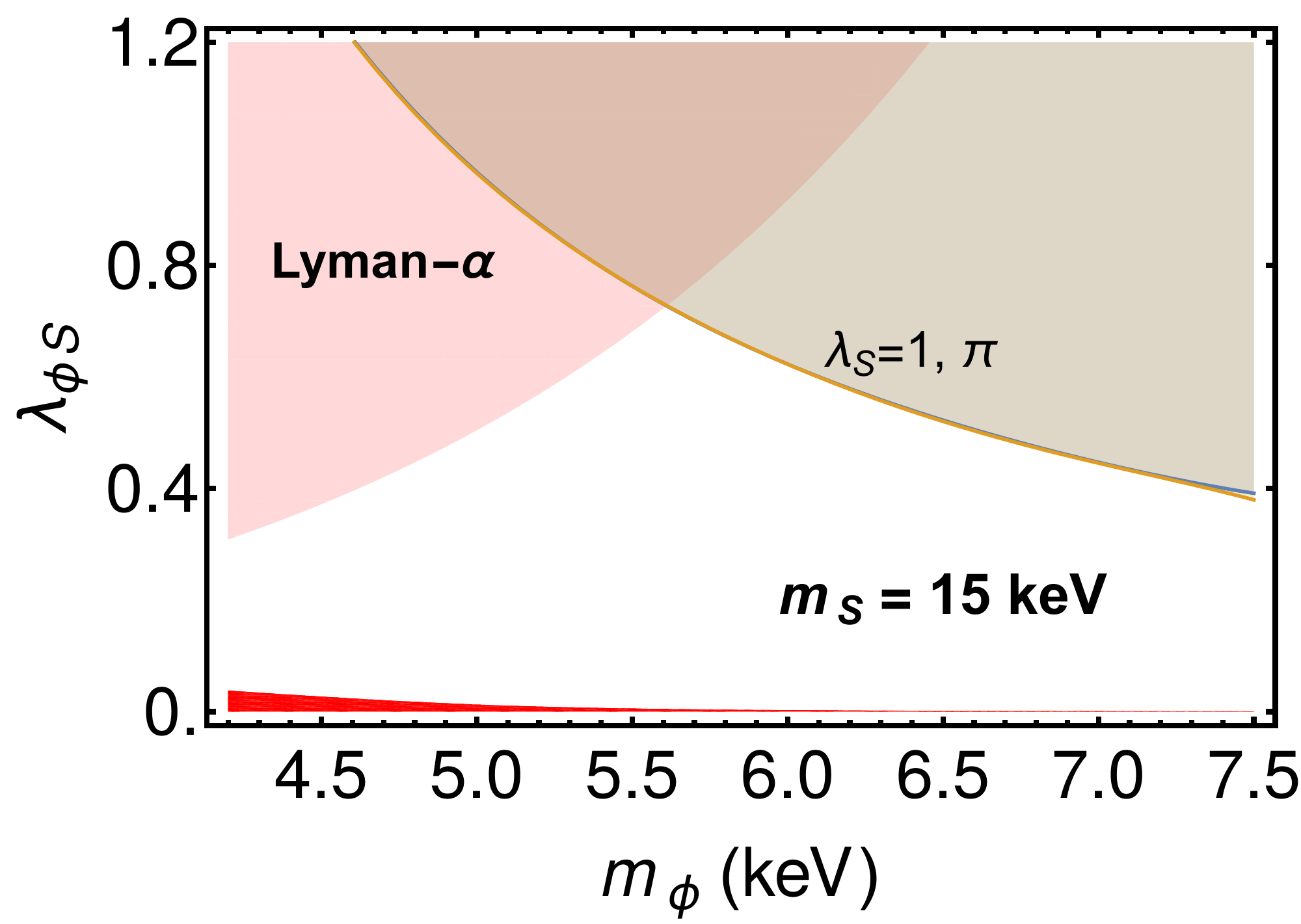}}~~
\subfloat[\label{sf:4to2_GTmSby2_mS15keV}]{
\includegraphics[scale=0.37]{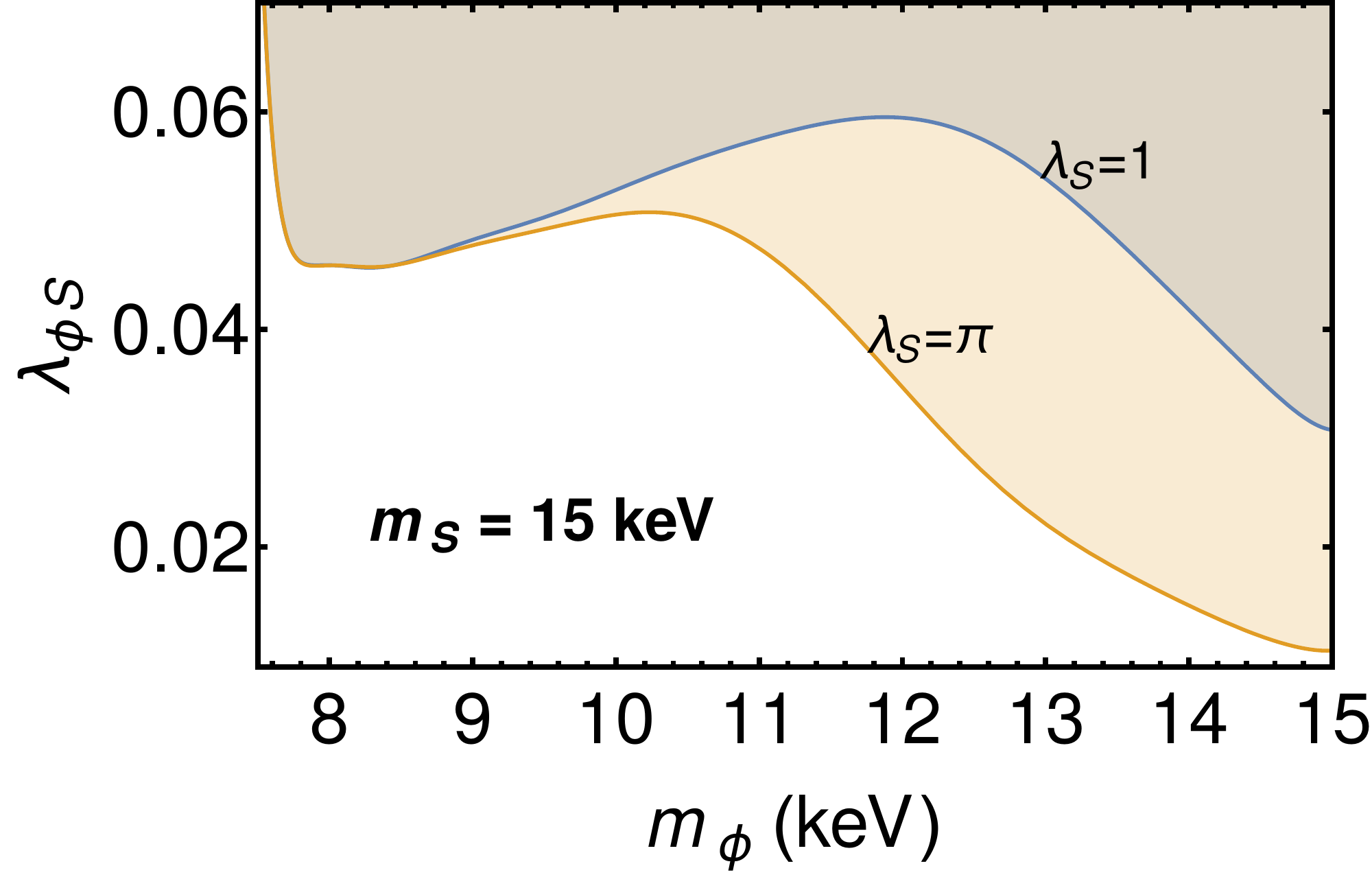}}\\
\subfloat[\label{sf:4to2_LTmSby2_mS30keV}]{
\includegraphics[scale=0.35]{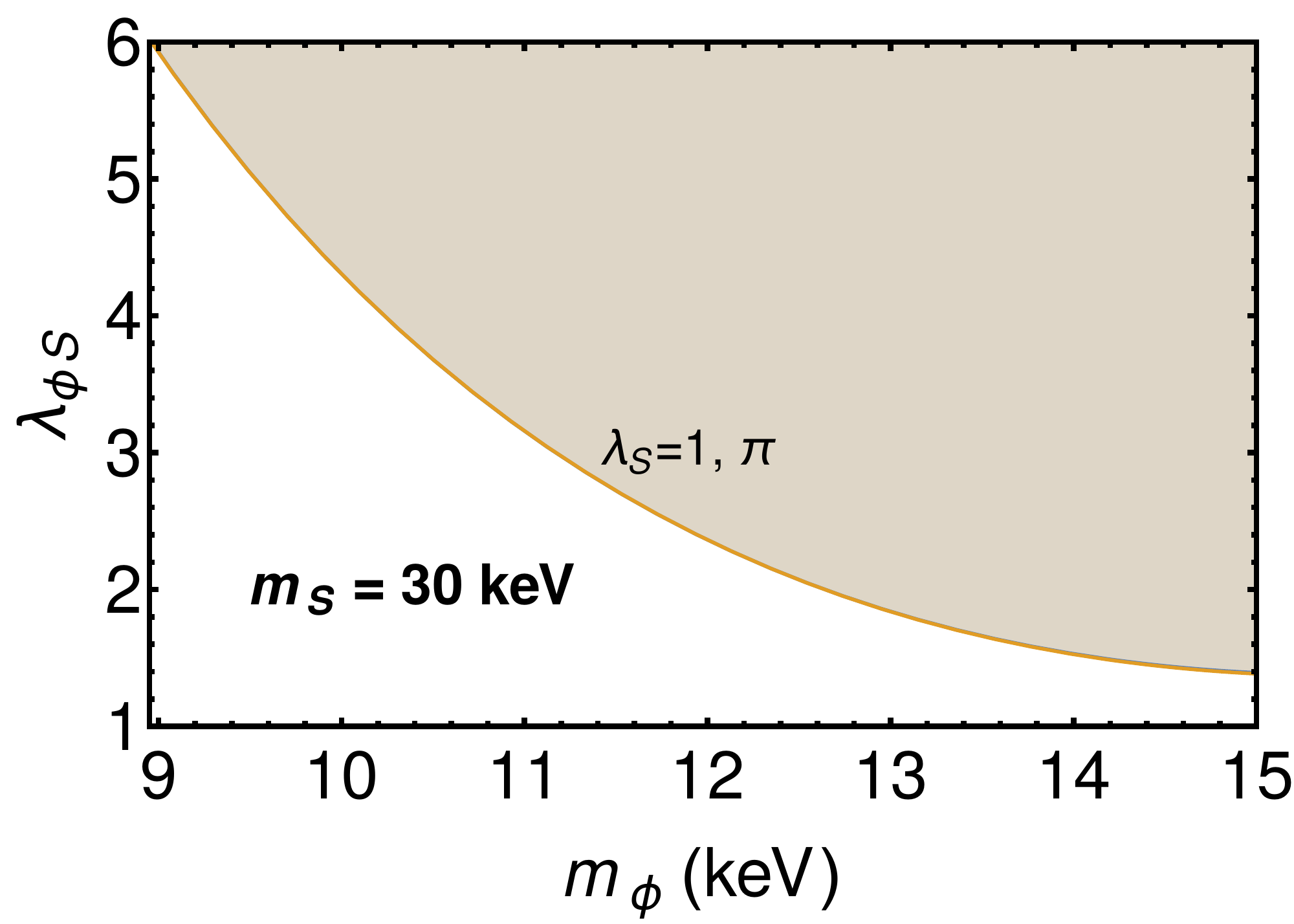}}~~
\subfloat[\label{sf:4to2_GTmSby2_mS30keV}]{
\includegraphics[scale=0.375]{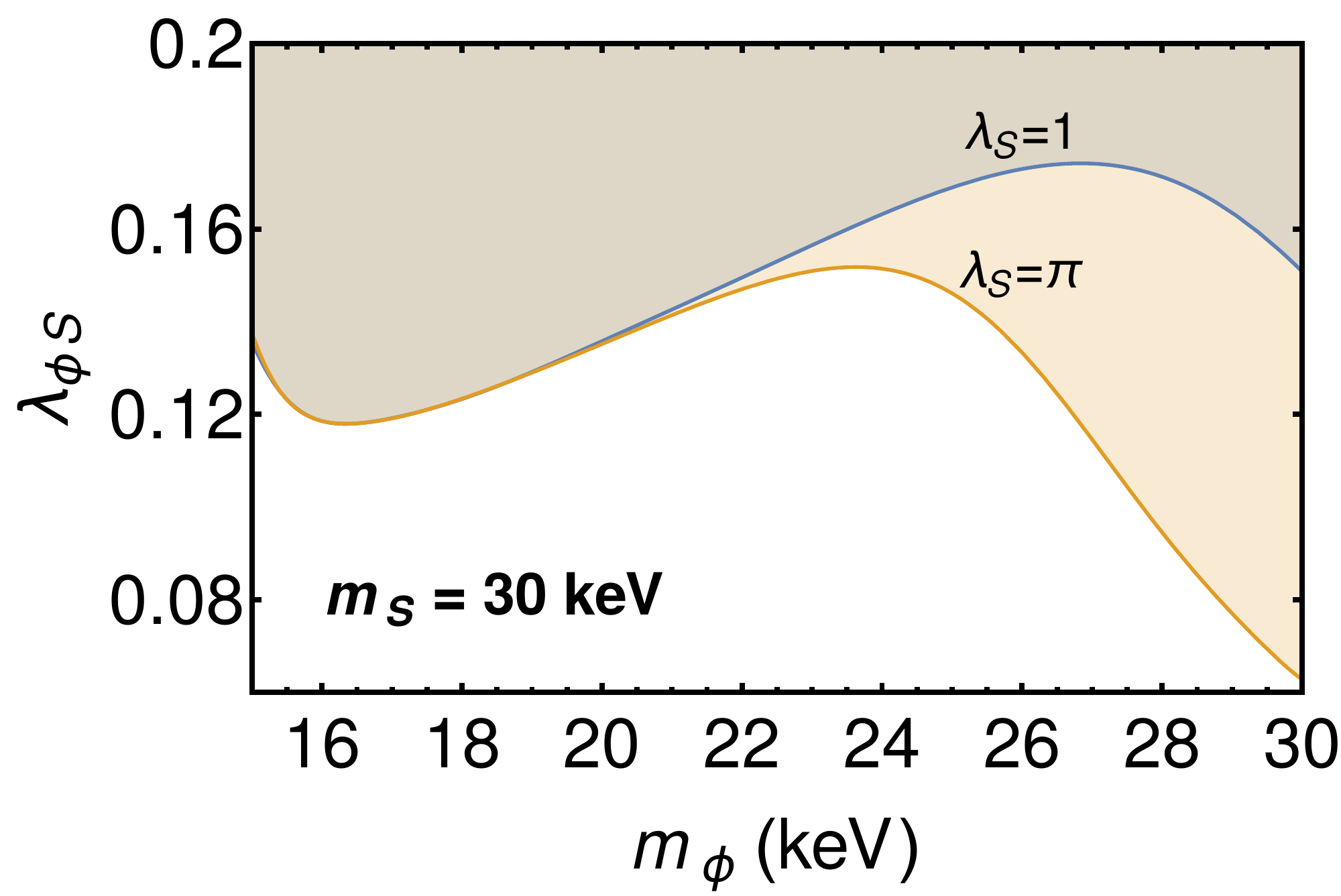}}
\caption{Allowed parameter space for the $4 \to 2$ processes in the $\lambda_{\phi S}$-$m_{\phi}$ plane for $\lambda_{S} = 1$ and $\pi$. (a) For $m_{\phi} < m_{S}/2$ with $m_{S} = 15$ keV. The light red region represents the Ly-$\alpha$ constraints, i.e., the region with $T_{f} < 1$ keV. The thin dark red region at the bottom shows the region of semi-relativistic freeze-out, i.e., where $x_{f} < 3$. The shaded region above the almost overlapping curves of $\lambda_{S} = 1$ and $\pi$ is the allowed region by relic density constraint. (b) For $m_{\phi} \in (m_{S}/2, m_{S})$ with $m_{S} = 15$ keV, the shaded region shows the relic density allowed region. (c) Allowed parameter space for $m_{\phi} < m_{S}/2$ with $m_{S} = 30$ keV and for (d) $m_{\phi} \in (m_{S}/2, m_{S})$ with $m_{S} = 30$ keV.}
\label{fig:4to2param}
\end{center}
\end{figure}

The dominant decay of a keV and MeV scale assisters considered in this paper, is to two photons via a $\tau$ loop as shown in figure~\ref{fig:syy}. The decay width is given by,
\begin{align}
 \Gamma \sim \frac{1}{16 \pi m_{S}} \left|\frac{\lambda_{\tau} \alpha_{\rm em} m^2_S }{6 \pi m_{\tau}}\right|^{2}.
\label{eq:decaywidth}
\end{align}
For $\lambda_{\tau} = 0.3$ the suppressed decay results in a life-time for the assisters $\sim 1$ seconds, that is safe from cosmological consequences.

Note that a keV scale DM has a relatively large free-streaming path of the order of a $\sim$ few Mpc, so their self-interaction cross section $\sigma_{\phi\phi \to \phi\phi} / m_{\phi}$ is constrained by several structure formation issues. A detailed discussion on such limits can be found in~\cite{Modak:2015npa} and references therein. A conservative limit on $\sigma_{\phi\phi \to \phi\phi} / m_{\phi} < 1$ cm$^{2}$g$^{-1}$ requires $\lambda_{\phi}$ to be of the order of  $10^{-6}$. Therefore, processes involving this coupling will have numerically insignificant effect on the relic density calculation. So we will neglect $\lambda_{\phi}$ in what follows. Interestingly, one may freely tune $\lambda_{\phi}$ so that $\sigma_{\phi\phi \to \phi\phi} / m_{\phi} \in (0.1,1)$ cm$^{2}$g$^{-1}$ \cite{Bernal:2015xba} to address galactic scale structure formation issues like {\it missing satellite}, {\it too-big-to-fail}, {\it core vs cusp} etc. problems without affecting the relic density discussion that follows. 
A viable region for DM through the assisted annihilation is obtained when both the DM and assister are at the keV scale. Assuming that $m_{\phi}$ is less than $m_S$, possible $4 \to 2$ annihilation process are shown in figure~\ref{fig:feyndia4to2}. The detailed calculation of  amplitudes of these processes, within the non-relativistic (NR) approximation, are given in appendix~\ref{appn:feyndia}. Note that a two-loop annihilation of $\phi$ via the channel $\phi \phi \to \gamma \gamma$ is possible as shown in figure~\ref{fig:phiann}. The estimate in the appendix \ref{appn:2to2} implies $\langle\sigma v\rangle \sim (\lambda_{\phi S}/10)^{2} (m_{\phi}/15~{\rm keV})^{2}\times 10^{-25}~\mbox{GeV}^{-2},$
having negligible numerical impact in the relic density calculation. Similarly $\phi \phi \to \gamma \gamma \gamma \gamma$, which also arises at two-loop, can be neglected.  
We numerically solve the relevant form of Boltzmann equation (given in eq.~\ref{eq:boltzy}) using correct annihilation cross sections to find out the region of parameter space allowed by relic density constraint from Planck data with $\Omega h^2 = 0.1161\pm 0.0028$~\cite{Ade:2013zuv}. Figure~\ref{fig:4to2param} depicts available parameter space for $m_{S}=15$ keV, $m_S=30$ keV.  We have checked that in the region  allowed by dark matter relic density,  the ratio of the mass of the dark matter to the freeze-out temperature,  $x_{f} > 3.$  This ensures that the dark matter is non-relativistic at the time of decoupling implying a  standard CDM scenario. We solve the Boltzmann equation semi-analytically to obtain $x_f$ as explained in appendix~\ref{appn:xf}. We impose conservative limit $T_f< 1$ keV which comes from the Lyman-$\alpha$ forest~\cite{Viel:2013apy}. The discontinuity in the allowed parameter space, which is evident between the figure~\ref{sf:4to2_LTmSby2_mS15keV} (\ref{sf:4to2_LTmSby2_mS30keV}) and \ref{sf:4to2_GTmSby2_mS15keV} (\ref{sf:4to2_GTmSby2_mS30keV}) is basically  because two sets of annihilation processes are operative in the two regime. One set of processes is kinematically allowed when $m_{\phi} < m_{S}/2$  while some other channels open up in the regime  $m_{S}/2 < m_{\phi} < m_{S}$ as shown in figure \ref{sf:4phito2S}. When $m_{\phi}< {m_S}/{2}$ freeze-out is achieved purely through the assisted annihilation/semi-annihilation processes, shown in figure~\ref{sf:3phiSto2S} and \ref{sf:2phi2Sto2S}.

\subsection{MeV scale dark matter}
\begin{figure}[t]
\begin{center}
\includegraphics[scale=0.7]{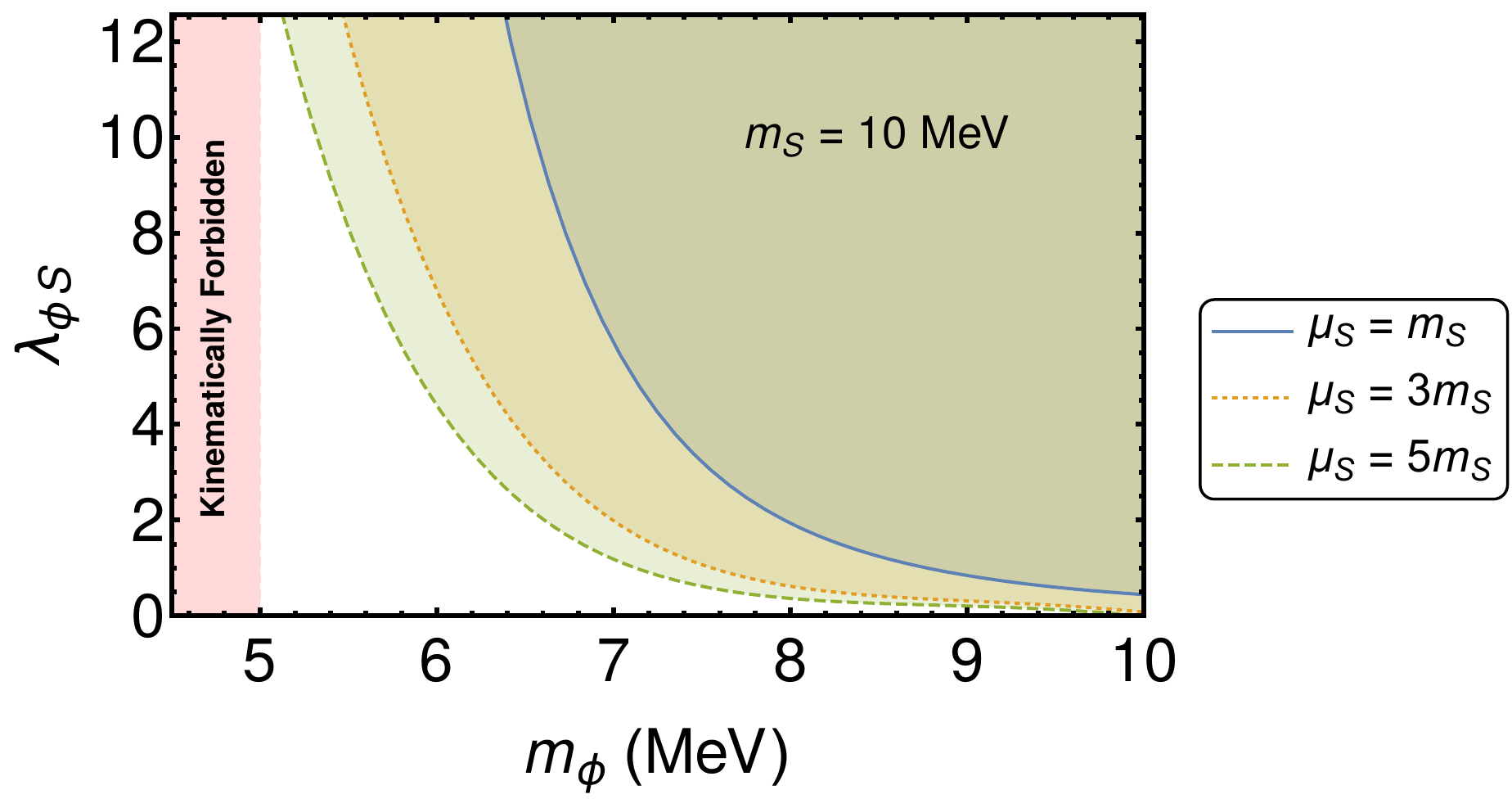}
\caption{Allowed parameter space, in the $\lambda_{\phi S}$-$m_{\phi}$ plane, satisfying the relic density criterion for $3\to 2$ processes with $m_{\phi} > m_{S}/2$ for $m_{S} = 10$ MeV. The light red band at the extreme left shows the region where the required $3\to 2$ process is kinematically forbidden. The shaded region shows the parameter space allowed from relic density constraints.}
\label{fig:10mev}
\end{center}
\end{figure}

The lightness of DM in the scenario of last section is due to large suppression factor of $4 \to 2$ annihilation cross section, as can be read off from eq.~\ref{eq:dim}. This cross section being suppressed by the DM and assister masses can lead to appreciable annihilation only if the spectrum is in keV scale. A heavier assister and DM mass spectrum is allowed if we incorporate $3 \to 2$ assisted annihilation channels. This can be easily achieved by modifying the Lagrangians described in eqs.~\ref{eq:lag} and \ref{eq:lagz2brk} by including the following self-interaction of the assister $S$,
\begin{align}
\label{eq:lag3to2}
\mathcal{L}_{3 \to 2} = \frac{\mu_S}{3!}S^3.
\end{align}
%
If we assume that the assister field does not get a vacuum expectation value the ratio of the  trilinear coupling in eq.~\ref{eq:lag3to2} to the assiter masses $m_{S}$ is bounded from above. Including metastability this limit tantamounts to $\mu_{S}/m_{S} \lesssim 7 $ \cite{Belanger:2012zr}. We will restrict ourselves to smaller values of this trilinear coupling.
In the present setup a $3 \to 2$ assisted annihilation channel as shown in figure~\ref{fig:feyndia3to2} opens up provided $m_{\phi} > m_{S}/2$. This can lead to a viable DM scenario in  the MeV scale as demonstrated in figure \ref{fig:10mev} where we show the allowed parameter space in the $\lambda_{\phi S}$-$m_{\phi}$ plane, satisfying the relic density criterion for $3\to 2$ assisted annihilation processes with for $m_{S} = 10$ MeV. The light red band at the extreme left shows the region where the required $3\to 2$ process is kinematically forbidden at tree-level. The shaded regions show the parameter space allowed from relic density constraints. The $4 \to 2$ assisted annihilation processes, discussed in the previous section, will remain simultaneously operative but will have the negligible numerical impact due to large suppression. The two-loop annihilation $\phi \phi \to \gamma \gamma$ will also be there but it will have negligible impact on the relic density calculation. As can be seen in the figure, smaller values of the DM-assiter coupling preferred by perturbativity up to high scale imply a more degenerate spectrum. We have checked that the region of interest in the parameter space does not have any issue with either semi-relativistic freeze-out or constraints from large scale structure formation. Again, enough freedom remains in the theory to tune the self-interaction to address structure formation issues at the galactic scale.

\section{Sketch of a vector boson assister model} 
\label{sec:vecAsst}
Instead of scalars one can also consider vector assisters. Below we give a brief sketch of such a scenario where the annihilation of scalar DM particles are assisted by vector bosons. 
Consider a scalar DM $\phi$ stabilised by a $\mathbb{Z}_{2}$ symmetry and charged under a hidden gauged $U(1)_{X}$  group  where $X_{\mu}$ is the corresponding gauge boson which will act as the assisters in this setup. Let this group be broken by a hidden Higgs mechanism or through an abelian Stueckelberg mechanism providing a mass for the gauge boson. Indeed one can arrange for the $U(1)_{X}$ symmetry to break in a way that leaves an unbroken discrete $\mathbb{Z}_{2}$ symmetry which can stabilise the DM. Further, we will assume that the $X_{\mu}$ field have a non-trivial coupling to some massive SM fermion(s). The assisters are assumed to decay to the SM states principally through this coupling. The effective Lagrangian for this setup can be written as,
\begin{align}
\mathcal{L}_{\rm vec} \supset \frac{1}{2}
                        \left(D_{\mu}\phi\right)^{2} + 
                        \frac12 m_{X}^{2} X_{\mu}X^{\mu} + 
                        \frac12 m_{\phi}^{2} |\phi|^{2} +
                       g_{X}^{\prime} \bar{f}\gamma_\mu f X^\mu,
\end{align}
where the covariant derivative is given by $D_{\mu} = \partial_{\mu} - i g_{X}X_{\mu}$ and the parameter $g_X'$ is the effective coupling of the boson $X_{\mu}$ with massive SM fermion, $f$. 
\begin{figure}[!htbp]
\begin{center}
\includegraphics[scale=0.5]{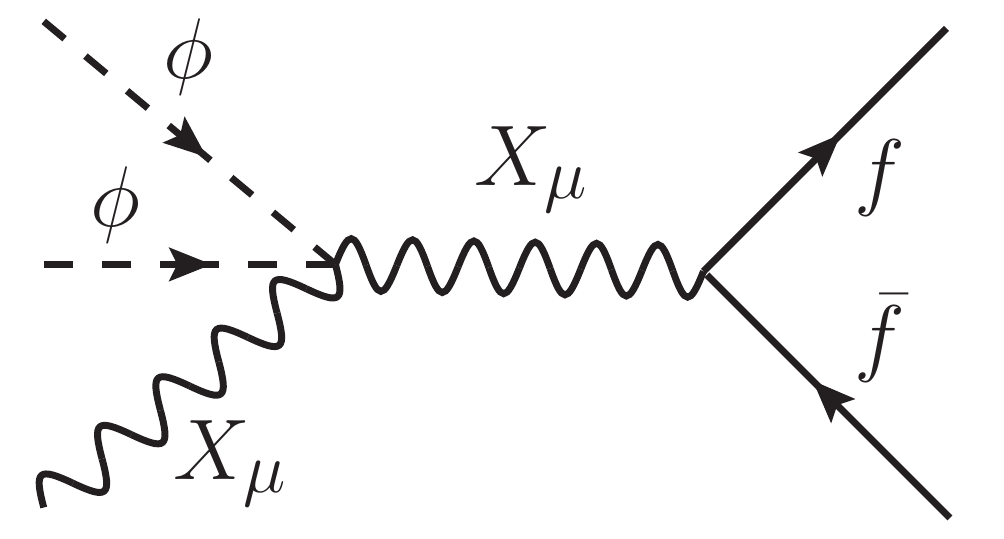}
\caption{Feynman diagram for a typical process with a vector assister.}
\label{fig:vecAsst}
\end{center}
\end{figure} 
If we assume $(m_{f} - m_{X}/2) < m_{\phi} < m_X $ then the annihilation of the DM states will proceed through figure~\ref{fig:vecAsst}. The coupling of the hidden massive gauge boson with the SM fermions are severely constrained from the electroweak precision test etc. Also there is the question of anomaly for this hidden gauge group. These are questions related to the details of the model and is beyond the scope of the present discussion.

\section{Summary and conclusions} 
\label{sec:concl}
In this paper we make a feasibility study for assisted annihilation to drive successful freeze-out of DM and account for its observed relic abundance in the Universe. Assisted annihilation may be relevant in the light dark matter scenarios when the DM is kinematically forbidden to annihilate to SM or SM-like states at tree-level. Such light dark matter can then be assisted by the assisters to annihilate and produce the required relic density. We find that for adequate annihilation a relative degeneracy in the mass of the assister and the DM is required. Both the non-standard topology of the annihilation process and the lightness of the DM makes the setup relatively insulated from direct or indirect detection experiments.
We have presented a simple model involving a scalar dark matter and a scalar assister that demonstrate the main features of this generic framework of assisted annihilation/semi-annihilation. We assume that the scalar assister separately couple to the SM states that allow the decay of the assisters keeping it in thermal equilibrium through out the freeze-out process. The scalar DM being the lightest state in the spectrum can only annihilate through processes involving more than two initial states including $4 \to 2$ and $3 \to 2$ assisted annihilation/semi-annihilation. We find that annihilation dominated by the $4 \to 2$ processes leads to keV scale DM while those involving $3 \to 2$ processes results in MeV scale DM. Generalisation of this framework to more involved setup is straightforward and an example with vector boson assisters is also sketched out.  
Such light DM having a large free streaming length in the early universe can significantly modify the structure formation at the scales of the $\sim$ few Mpc. Astrophysical observations like constraints from Lyman-$\alpha$ forest provide the  most relevant tests for these models. The coupling of the assisters with the SM states are constrained from various collider and precision observables. Searches for these assisters at colliders or  low scale precision observables may provide a complimentary strategy to constrain this framework. For example in the model presented in section \ref{sec:scalarAsst}, the assisters may be produced in a future photon-photon  collider \cite{Ginzburg:1981vm, Ginzburg:1982yr}. Also they will leave their imprint in observables like $(g-2)_\tau.$ Though it is a challenge to  estimate the anomalous magnetic moment of tau owing to its short lifetime,  several proposal for improved measurement  do exist in the literature \cite{Eidelman:2007sb}.  A model dependent study of dedicated searches for the assisters  is an interesting subject for further investigation.

\acknowledgments
We thank Andrew Spray, Somnath Bharadwaj, and Arindam Chatterjee for useful discussions and comments. UKD acknowledges the support from Department of Science and Technology, Government of India under the fellowship reference number PDF/2016/001087 (SERB National Post-Doctoral Fellowship). TNM thanks MHRD, Government of India for fellowship. TSR acknowledges the ISIRD Grant, IIT Kharagpur, India.


\appendix
\section{Feynman diagrams and amplitudes for relevant processes}
\label{appn:feyndia}
In this appendix we present the details of the calculation of the amplitudes and Feynman diagrams for the relevant $3\to 2$ and $4\to 2$ processes in the ambit of the model presented in section~\ref{sec:scalarAsst}. All these amplitudes are in the non-relativistic approximation which implies the initial state particles are of zero momentum, i.e., we take $p^{\mu}_{\rm initial} = (m_{\rm initial},\vec{0})$. We also present an estimate of the two-loop $\phi \phi \to \gamma \gamma$ cross section.

\subsection[$4\to 2$ processes:]{\boldmath{$4\to 2$} processes:}
There are three relevant $4\to 2$ processes, namely, $\phi \phi \phi \phi \to SS$, $\phi \phi \phi S \to \phi S$ and $\phi \phi S S \to \phi S$. The corresponding Feynman diagrams are shown below.

\begin{figure}[!htbp]
\begin{center}
\subfloat[\label{sf:4phito2S}]{
\includegraphics[scale=0.4]{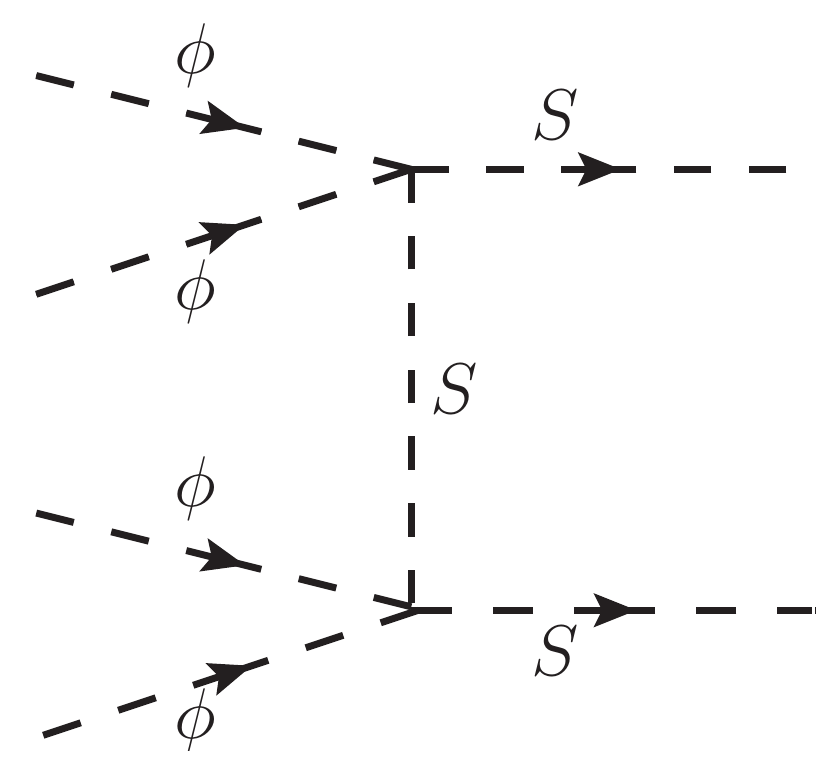}}~~~
\subfloat[\label{sf:3phiSto2S}]{
\includegraphics[scale=0.35]{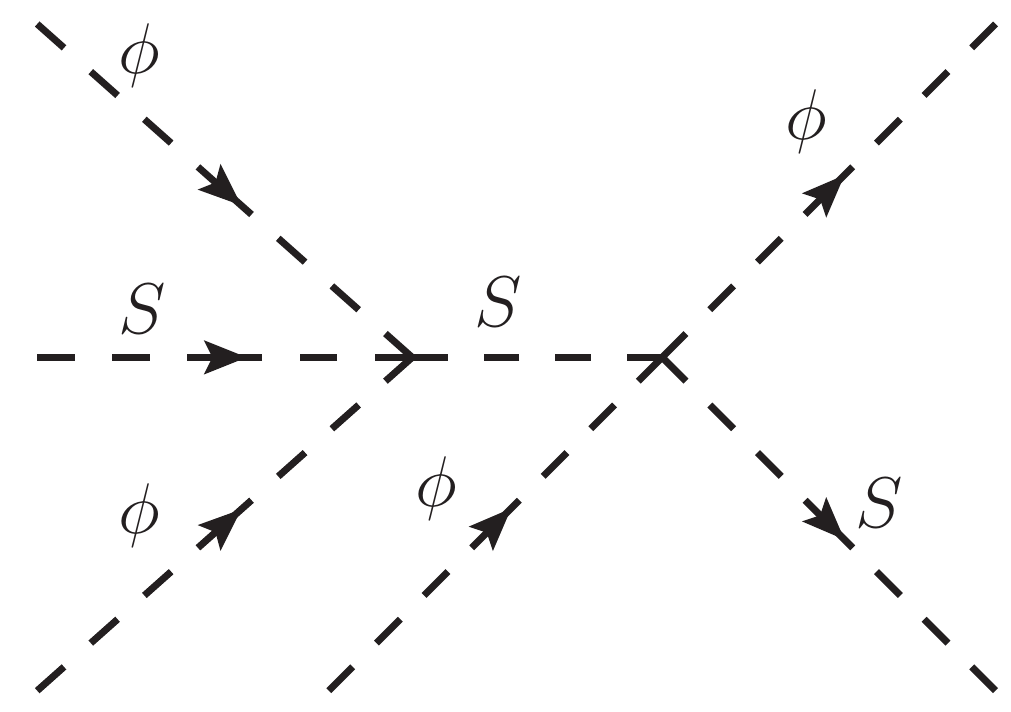}~~~
\includegraphics[scale=0.4]{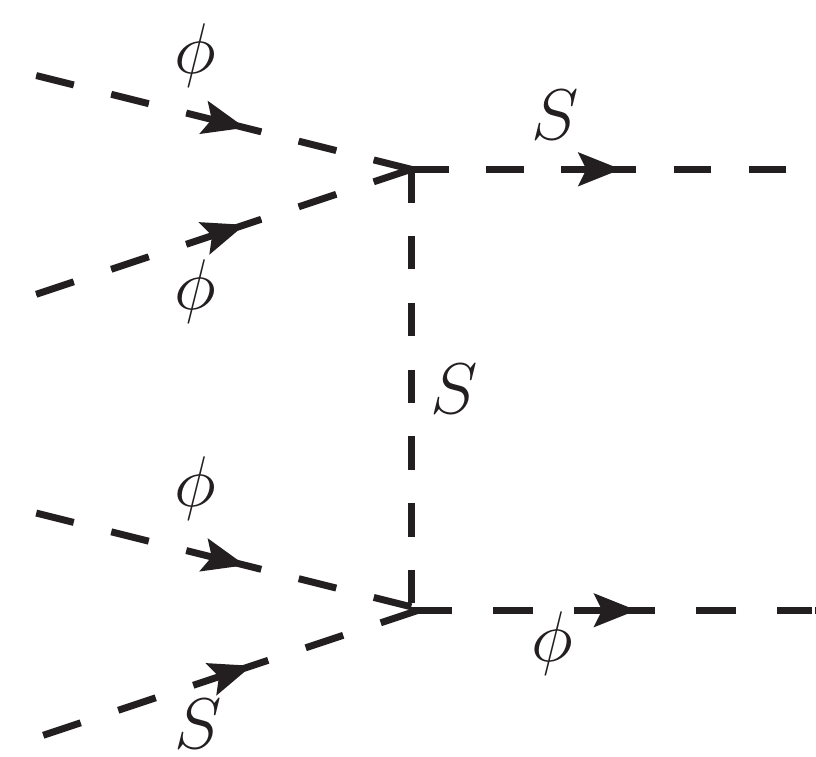}}\\
\subfloat[\label{sf:2phi2Sto2S}]{
\includegraphics[scale=0.4]{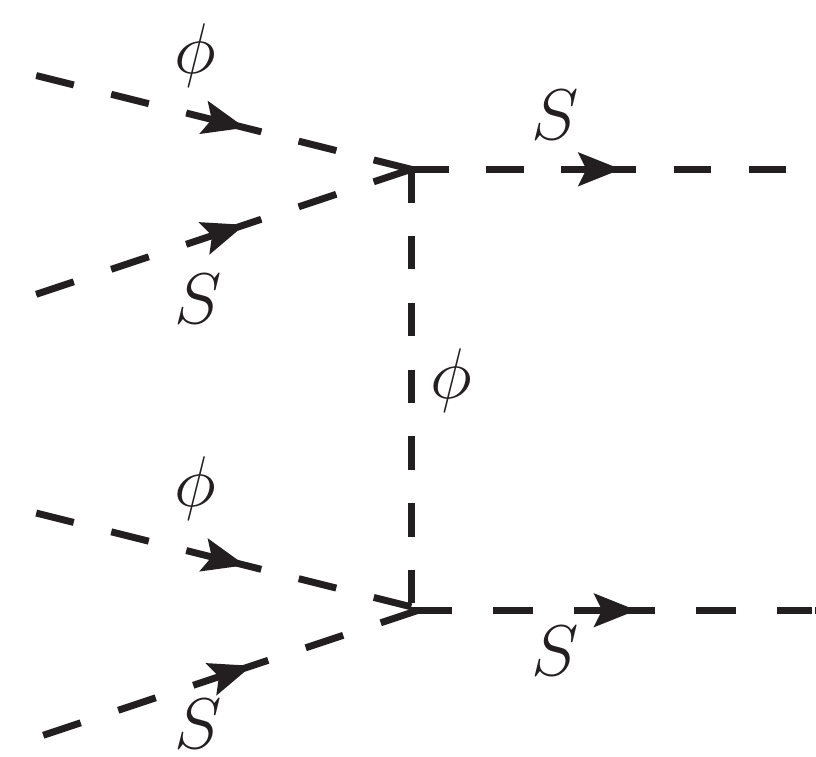}~~
\includegraphics[scale=0.4]{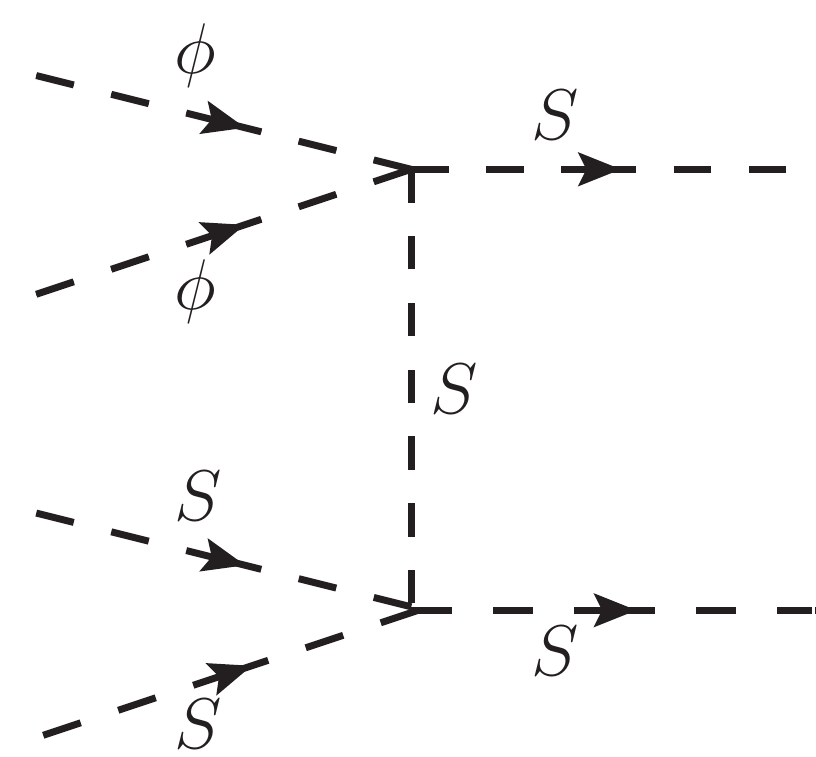}~~
\includegraphics[scale=0.35]{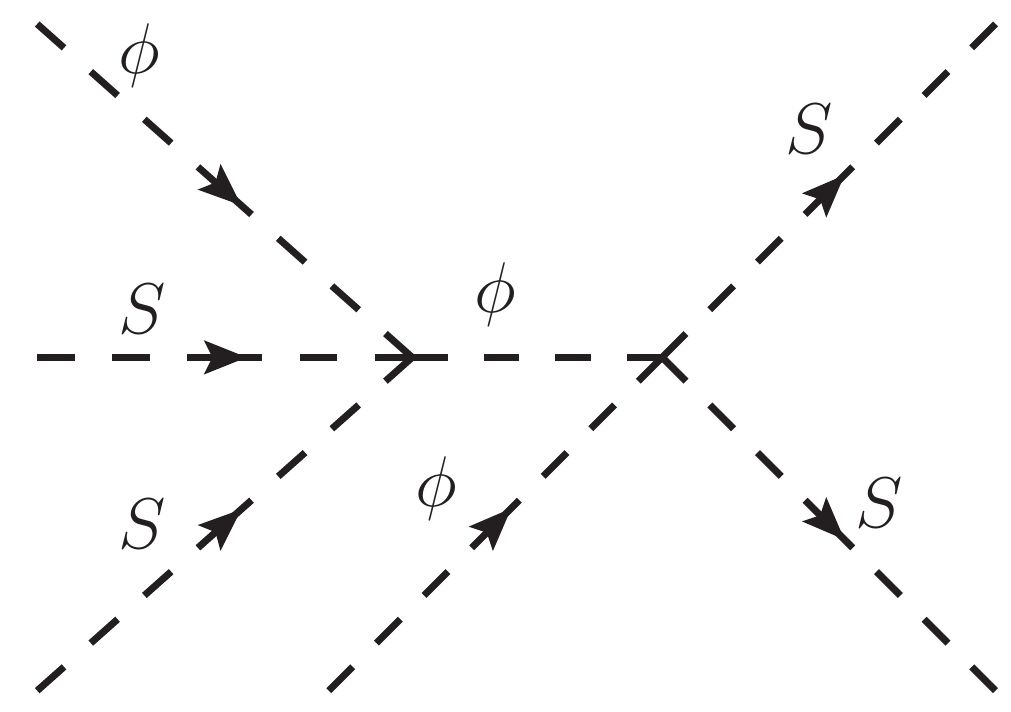}~~
\includegraphics[scale=0.35]{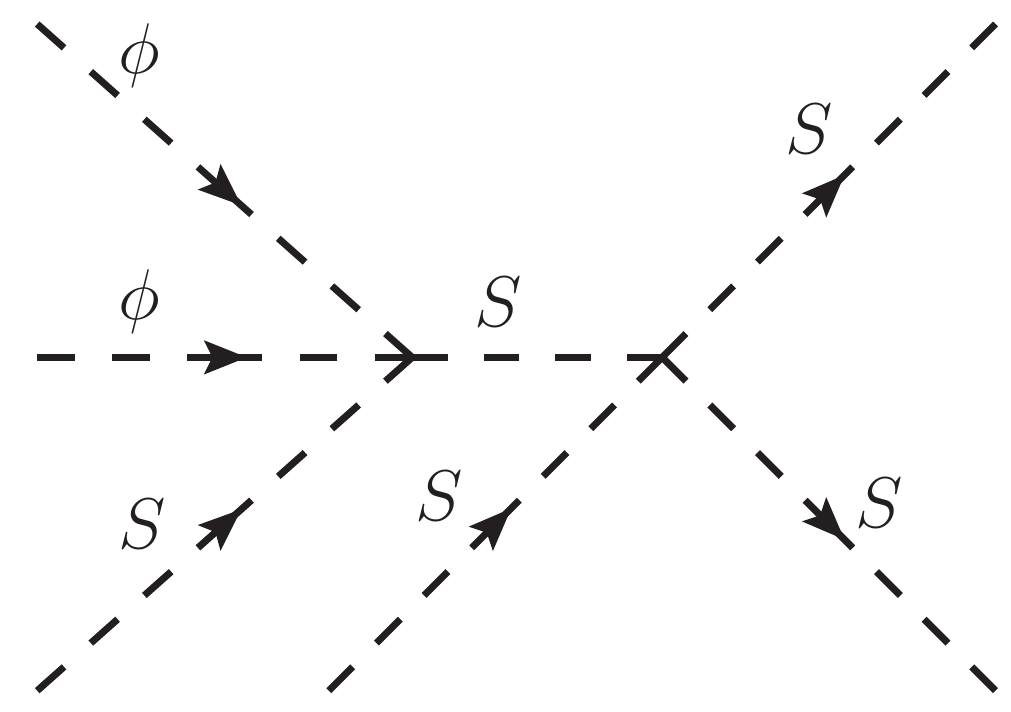}}
\caption{Feynman diagrams for the $4 \to 2$ processes.}
\label{fig:feyndia4to2}
\end{center}
\end{figure}

The amplitude for these processes in the non-relativistic approximation are given by,
\begin{subequations}
\begin{align}
\mbox{figure 8a:}~~\lvert \mathcal{M}_{\phi \phi \phi \phi 
                  \to S S}\rvert^{2}_{\rm NR} &=  
                    \frac{\lambda_{\phi S}^4}{16 m^4_{\phi}},\\
\mbox{figure 8b:}~~\lvert \mathcal{M}_{\phi \phi \phi S 
                  \to \phi S}\rvert^{2}_{\rm NR} &= 
                  \lambda_{\phi S}^4 
                  \Bigg(
                   \frac{3m_{\phi} + 
                    m_S}{4 m^2_{\phi} (3m_{\phi} + m_S) - 
                    2 m_{\phi} (3m_{\phi} + m_S)^2 + 
                    2 m_{\phi}(m^2_{\phi} - m^2_S)} \notag \\
                    & \qquad \qquad \qquad \qquad + 
                    \frac{1}{4 m^2_{\phi} + 4 m_{\phi} m_S} 
                   \Bigg)^{2},\\
\mbox{figure 8c:}~~\lvert \mathcal{M}_{\phi \phi S S 
                  \to S S}\rvert^{2}_{\rm NR} &= 
                  \lambda_{\phi S}^2
                  \Bigg(
                   \frac{\lambda_S}{4 m_{\phi}m_S} +
                   \frac{2\lambda_{\phi S} - \lambda_S}
                    {4m_{\phi}^2+4 m_{\phi}m_S} - 
                    \frac{\lambda_{\phi S}}{4m_{S}^2 + 
                    4 m_{\phi} m_S}
                  \Bigg)^{2}\,.                   
\end{align}
\end{subequations}

\subsection[$3\to 2$ processes:]{\boldmath{$3\to 2$} processes:}
The relevant $3\to 2$ process is $\phi \phi S \to SS$. The Feynman diagrams for this process are shown below.
\begin{figure}[!htbp]
\begin{center}
\includegraphics[scale=0.4]{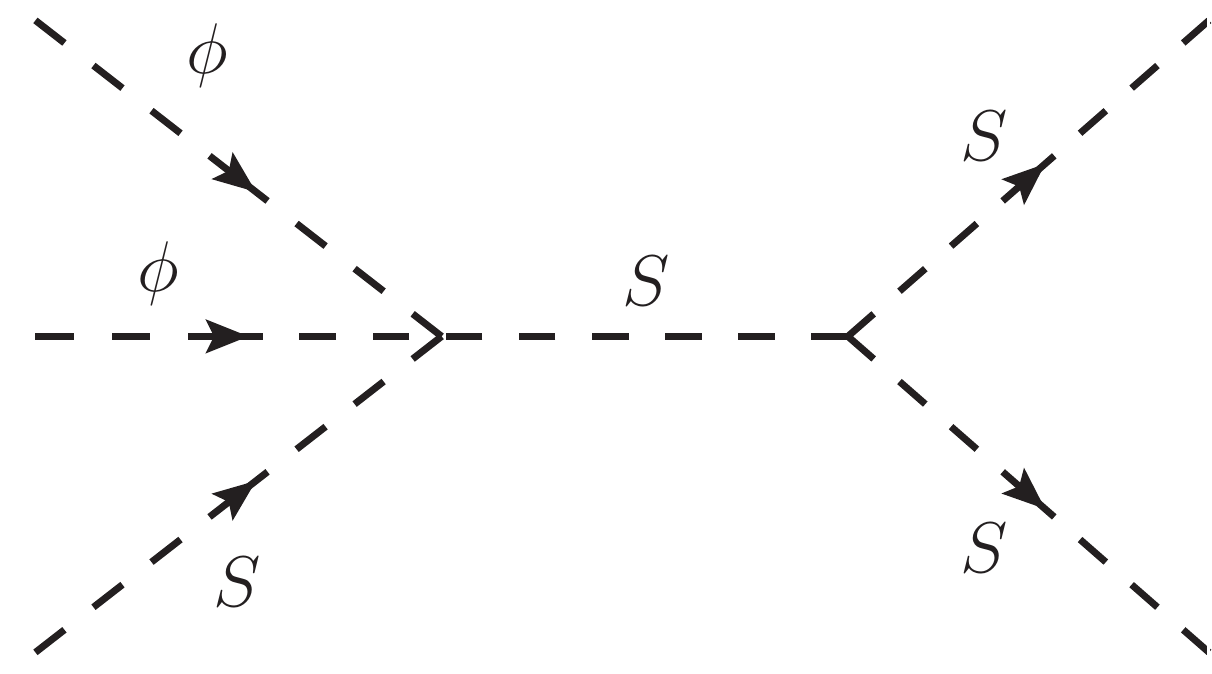}~~~~~
\includegraphics[scale=0.4]{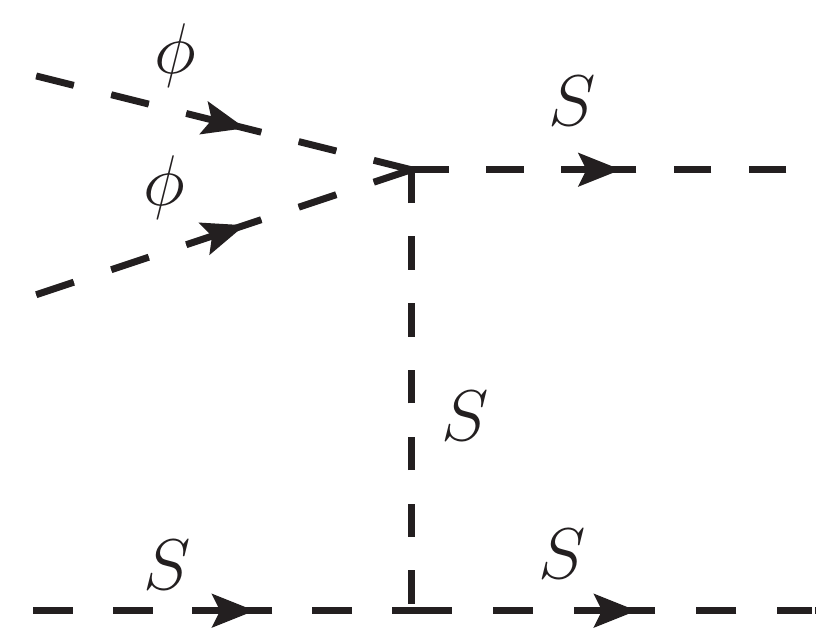}
\caption{Feynman diagrams for $\phi \phi S \to SS$.}
\label{fig:feyndia3to2}
\end{center}
\end{figure} 

The amplitude for this process in the non-relativistic approximation is given by,
\begin{align}
\lvert \mathcal{M}_{\phi \phi S \to S S}\rvert^{2}_{\rm NR}=
      \frac{\lambda_{\phi S}^{2} \mu^2}{m^2_S}
      \left(\frac{m_S}{(2m_{\phi} + m_S)^2 - m^2_S} - 
      \frac{1}{m_{\phi}} \right)^2.
\end{align}

\subsection[Loop-induced $2\to 2$ annihilation:]{Loop-induced \boldmath{$2\to 2$} annihilation:}
\label{appn:2to2}
\begin{figure}[!htbp]
\begin{center}
\includegraphics[scale=0.36]{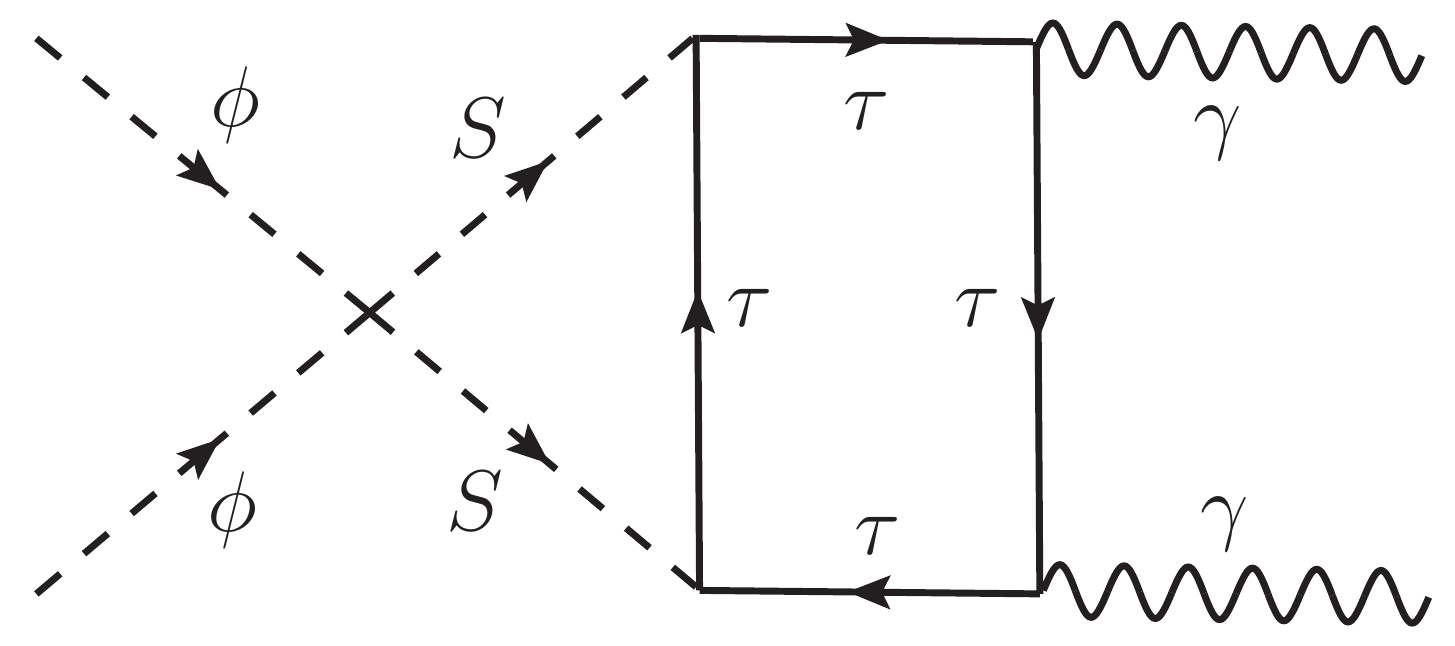}
\caption{Feynman diagrams for loop-induced annihilation of DM particle $\phi$.}
\label{fig:phiann}
\end{center}
\end{figure} 
A naive dimensional estimate of the thermally averaged cross section, in the limit $m_{\phi}\sim m_{S} \ll m_{\tau}$, yields
\begin{align*}
 \langle \sigma v\rangle \big|_{\rm NR} \sim \frac{1}{128\pi^{2}m_{\phi}^{2}}
                               \left|\frac{\lambda_{\phi S}}{16\pi^{2}}
                               \frac{\alpha_{\rm em}}{4\pi}\lambda_{\tau}^{2}
                               \frac{m_{\phi}^{2}}{m_{\tau}^{2}}\right|^{2}.
\end{align*}

\section{Calculation of freeze-out temperature}
\label{appn:xf}
Here we outline the general procedure of calculating the freeze-out temperature $x_{f}$ corresponding to the $4\to 2$ assisted annihilation/semi-annihilation processes. One can easily follow this procedure to obtain $x_{f}$ for $3\to 2$ processes also. The relevant Boltzmann equation in this case is given by,
\begin{align}
&\frac{dY}{dx} = -\frac{1}{x^{8}}
                 \left[
                 F_{\phi \phi \phi \phi \to SS}
                 \left(Y^{4} - Y^{4}_{\rm eq}\right) + 
                 F_{\phi \phi \phi S \to \phi S}
                 \left(Y^{3}Y_{\rm eq} - Y Y^{3}_{\rm eq}\right) + 
                 F_{\phi \phi S S \to SS}
                 \left(Y^{2}Y^{2}_{\rm eq} - Y^{4}_{\rm eq}\right)  
                 \right], \\
&\mbox{where}~~~ F_{X} = \frac{s^{3}(m_{\phi})}{H(m_{\phi})}
          N_{\rm Bolt}(q)\langle \sigma v^{3} \rangle_{X},
\end{align}
and $N_{\rm Bolt}(q)$ is given by eq.~\ref{seq:nbolt} with $q$ being the number of assister $S$ in the initial state. 
Then, in terms of $\Delta = Y-Y_{\rm eq}$ the Boltzmann equation becomes,
\begin{align}
\Delta^{\prime} + Y_{\rm eq}^{\prime} = -\frac{1}{x^{8}}
               \left(\Delta + 2Y_{\rm eq}\right)\Delta
               \bigg[F_{\phi \phi \phi \phi \to SS}
               \left\lbrace \left(\Delta + Y_{\rm eq}\right)^{2} + 
                 Y_{\rm eq}^{2}\right\rbrace 
                 &+ F_{\phi \phi \phi S \to \phi S}
                    Y_{\rm eq}(\Delta + Y_{\rm eq}) \notag \\
                  &+ F_{\phi \phi S S \to S S} Y_{\rm eq}^{2} 
                 \bigg]
\label{eq:DelBoltz}
\end{align}
Now, with standard assumptions, from the above equation one can write,
\begin{align}
\Delta = \frac{x^{8}}{2 Y_{\rm eq}^{2}\tilde{F}}
         \left(1 - \frac{3}{2x}\right),
\label{eq:Del}
\end{align}
where $\tilde{F} = 2F_{\phi \phi \phi \phi \to SS} + F_{\phi \phi \phi S \to \phi S} + F_{\phi \phi S S \to S S}$. The freeze-out is defined as $\Delta(x_{f}) = c Y_{\rm eq}(x_{f})$, and noting that $Y_{\rm eq} = a e^{-x} x^{3/2}$ (where $a=0.145/g_{\ast}$) one can write using eqs.~\ref{eq:DelBoltz} and \ref{eq:Del}, the equation for $x_{f}$ as
\begin{align}
\sqrt{x_{f}} e^{3 x_{f}}\left(x_{f}^{3} - 
         \frac{3}{2} x_{f}^{2}\right) = 2\tilde{F} a^{3}c.
\end{align}
This equation can be solved to obtain the numerical value of the freeze-out temperature $T_{f}(=m_{\phi}/x_{f})$. In our calculation we have taken $c=1$.

\bibliographystyle{JHEP}
\bibliography{lightdm_ref.bib}

\end{document}